\documentclass[preprint, times]{elsarticle}
\usepackage{amssymb,amsthm,amsmath,float, booktabs}

\usepackage[total={6.5in,8.75in}, top=1.2in, left=0.9in, includefoot]{geometry}
\usepackage{graphicx}
\usepackage{epstopdf}
\usepackage[pdftex,bookmarks, colorlinks, citecolor =blue, breaklinks]{hyperref}

\newcommand{\Eq}[1]{(\ref{eq:#1})}

\newcommand{\Sec}[1]{\S \ref{sec:#1}}
\newcommand{\Fig}[1]{Fig.~\ref{fig:#1}}
\newcommand{\Figs}[2]{Figs.~\ref{fig:#1} and \ref{fig:#2}}
\newcommand{\Tbl}[1]{Table~\ref{tbl:#1}}

\newcommand{\InsertFig}[4]
{\begin{figure}[h!t]
       \centerline{
         \includegraphics[width=#4]{./figures/#1}
       }
       \caption{{\footnotesize  #2}
       \label{fig:#3}}
\end{figure}}

\newcommand{\InsertFigTwo}[5] {
\begin{figure}[ht]
       \centerline{
         \includegraphics[width=#5]{./figures/#1}
         \hskip 0.5in
         \includegraphics[width=#5]{./figures/#2}
       }
       \caption{{\footnotesize  #3}
       \label{fig:#4}}
\end{figure}}

\newcommand{\bR}{{\mathbb{ R}}}

\newcommand{\bT}{{\mathbb{ T}}}
\newcommand{\bZ}{{\mathbb{ Z}}}
\newcommand{\bN}{{\mathbb{ N}}}

\newcommand{\cL}{{\cal L}}

\newcommand{\cR}{{\cal R}}
\newcommand{\cT}{{\cal T}}

\newcommand{\eps}{\varepsilon}

\newcommand{\med}{\mathop{\rm median}}

\newcommand{\rank}[1]{{\mathop{\rm rank}({#1})}}

\theoremstyle{definition}

\newcommand{\beq}[1]{\begin{equation}\label{eq:#1}}
\newcommand{\eeq}{\end{equation}}

\newcommand{\bsplit}[1]{\begin{equation}\label{eq:#1}\begin{split}}
\newcommand{\esplit}{\end{split}\end{equation}}

\newenvironment{example}[1][]
  {
	\setlength \leftmargini {1.0em}		
	\setlength \topsep {0.5em}			
	\begin{quote}
	{\it Example#1} }
	{\end{quote}
  }
\newcommand{\bexam}[1][:]{\begin{example}[#1]}
\newcommand{\eexam}{\end{example}}

\begin{document}
\begin{frontmatter}

\title{The Destruction of Tori in Volume-Preserving Maps}
\author[cub]{J.D.~Meiss\corref{cor}\fnref{fn2}}
\ead{James.Meiss@colorado.edu}

\fntext[fn2]{JDM acknowledges support from NSF grant DMS-0707659, as well as many helpful discussions with Holger Dullin, Robert Easton, Brock Mosovsky and Adam Fox.}
\address[cub]{Department of Applied Mathematics, UCB 526, University of Colorado, Boulder, CO, 80309-0526, USA}

\date{\today}

\begin{abstract}
Invariant tori are prominent features of symplectic and volume preserving maps. From the point of view of chaotic transport the most relevant tori are those that are barriers, and thus have codimension one. For an $n$-dimensional volume-preserving map, such tori are prevalent when the map is nearly ``integrable," in the sense of having one action and $n-1$ angle variables. As the map is perturbed, numerical studies show that the originally connected image of the frequency map acquires gaps due to resonances and domains of nonconvergence due to chaos. We present examples of a three-dimensional, generalized standard map for which there is a critical perturbation size, $\eps_c$, above which there are no tori. Numerical investigations to find the ``last invariant torus" reveal some similarities to the behavior found by Greene near a critical invariant circle for area preserving maps: the crossing time through the newly destroyed torus appears to have a power law singularity at $\eps_c$, and the local phase space near the critical torus contains many high-order resonances. 


\end{abstract}

\begin{keyword}
	volume-preserving \sep incompressible \sep invariant tori \sep rotation vectors
\MSC{34C37, 37C29, 37J45, 70H09}
\end{keyword}
  
\end{frontmatter}

\section{Introduction}
Volume preserving maps are appropriate models for many systems including fluid flows 
\cite{Cartwright96, Meleshko99, Sotiropoulos01, Rodrigo03, Speetjens04, Anderson06, Mullowney08}, granular mixers \cite{Meier07}, magnetic field line flows 
\cite{Thyagaraja85, Greene93, Bazzani98}, and even the motion of comets perturbed by a planet on an elliptical orbit \cite{Liu94}.
Volume-preserving dynamics has some similarities to symplectic dynamics; however, though every symplectic map is volume preserving, the converse is only true in two-dimensions.  

This paper is concerned with the effects of perturbation and resonance on invariant tori; such tori are especially common in the integrable case.
In the context of Hamiltonian systems and symplectic maps, integrability is synonymous with Liouville's definition: a $d$-degree of freedom system is integrable when it has a set of $d$, almost-everywhere independent, involutory invariants. The involution property of the invariants implies that they also generate an Abelian group of symmetries that preserve the invariants, and therefore that the (compact) integral manifolds are tori. An integrable $2d$-dimensional symplectic map can be written (at least locally) in terms of angle-action variables $(\theta,J) \in \bT^d \times \bR^d$ as
\bsplit{integrable}
	\theta' &= \theta + \Omega(J) \;, \\
	J' &= J\;,
\end{split}\end{equation}
where we will take $\bT^d \equiv \bR^d/\bZ^d$  \cite{Veselov91}.
The concept of integrability is perhaps less well-formulated for the volume-preserving case. However, it seems quite natural to use Bogoyavlenskij's concept of \emph{broad integrability} \cite{Bogoy98,Fasso02}. Roughly speaking, a system of $n$ ODEs is broadly integrable if it has $k$ independent invariants and $d$ commuting symmetries that preserve the invariants, where $k+d = n$.
We propose a similar definition for a map---the integrable case corresponds to the form \Eq{integrable} as well, but now $(\theta,J) \in \bT^{d} \times \bR^k$  \cite{Lomeli11a}.

It is natural to study \Eq{integrable} on the universal cover, letting $(x,z) \in\bR^d \times \bR^k$, so that $(\theta, J) = (x \mod 1, z)$. We study a simple class of perturbations of \Eq{integrable}, which for the lift becomes
\bsplit{perturbed}
		x' &= x + \Omega(z') \;, \\
		z' &= z - \eps g(x) \;,
\end{split}\end{equation}
where the ``force," $g$, is periodic, i.e., $g(x+m) = g(x)$ for any $m \in \bZ^d$. Since $\Omega$ is evaluated at $z'$ in \Eq{perturbed}, this map is a volume-preserving diffeomorphism for any smooth functions $\Omega$ and $g$.\footnote
	{When $d = k$, \Eq{perturbed} is symplectic with the form $dx \wedge dz_i$ only if $D\Omega$ and $Dg$ are symmetric matrices.}
It is \emph{exact} volume preserving when $g$ has zero average---or equivalently the zero Fourier component of $g$ vanishes \cite{Lomeli09a}. Since this is the only case for which \Eq{perturbed} can have rotational tori---that is, tori homotopic to the tori of \Eq{integrable}---that are invariant, we will make this assumption.

The rotation vector (or frequency) map:
\beq{freqMap}
	\Omega: \bR^k \to \bR^d
\eeq
plays an especially important role in the dynamics of \Eq{integrable}. For the integrable map, the forward orbit $\{(x_t,z_t): t\in \bN\}$ of each initial condition $(x_0,z_0)$ has a rotation vector
\beq{rotationNumber}
	\omega(x_0,z_0) = \lim_{T \to \infty} \frac{x_T-x_0}{T}
\eeq
given by $\Omega(z_0)$. If $\Omega(z_0)$ is incommensurate, see \Sec{resonances}, the orbit densely covers a $d$-dimensional torus; by contrast, when the rotation vector is resonant an orbit densely covers one or more lower dimensional tori. 

When the map \Eq{integrable} is perturbed, many of its $d$-tori are immediately destroyed; however, KAM theory implies that there will still be a large set of invariant tori if the perturbation is small enough and smooth enough and the frequency satisfies a nondegeneracy condition. This is rigorously true for $k=d$ when the perturbed map is exact symplectic and  satisfies a H\"older condition (i.e. is $C^{3+h}$ for some h>0), and $\Omega$ satisfies a nondegeneracy condition such as the \emph{twist condition} 
\beq{Twist}
	\det{D\Omega} \ge c > 0 \;,
\eeq
see, e.g., \cite{Poshel01, delaLlave01}.

Of course, when $k<d$, the number of actions is smaller than the number of angles and the matrix $D\Omega$ is no longer square. The ``nicest" case corresponds to $\rank{D\Omega} = k$ implying that the image of $\Omega$ is an immersed $k$-dimensional submanifold.

Maps of the form \Eq{integrable} with $k=1$, so-called \emph{one-action} maps \cite{Piro88}, have codimension-one invariant tori. KAM theory implies that codimension-one tori are robust features of nearly-integrable, analytic one-action maps \cite{Cheng90b, Xia92}. These theorems assume that $\Omega \in C^{d+1}$ and satisfies a nondegeneracy condition of the form
\beq{Wronskian}
	\det (D\Omega, D^2\Omega, \ldots D^{d}\Omega) \ge c > 0 \;,
\eeq
similar to that used by R\"ussmann \cite{Russmann01, Sevryuk06}.

By contrast, when $1 < k < d$ the invariant $d-k$ dimensional tori of \Eq{integrable} need not be as robust. For example when there are two actions and one angle, almost all of the one-dimensional tori can apparently be immediately destroyed even under a smooth perturbation \cite{Mezic01}.

Though codimension-one tori are commonly observed in near-integrable, one-action maps, they are often destroyed by resonant bifurcations as the perturbation grows. The nature of these bifurcations is strongly influenced by the form of the frequency map. Even when the map satisfies \Eq{Wronskian} perturbations of \Eq{integrable} can have many of the features of symplectic maps that do not satisfy the twist condition \cite{Dullin10a}, see \Sec{Tangency}. In \Sec{lastTorus}, we will investigate which of these persistent tori is most robust.

\section{Frequency Maps and Resonance}\label{sec:resonances}
A rotation vector $\omega \in \bR^d$ is ``resonant" when there exists an $(m,n) \in \bZ^d \times \bZ \setminus \{0, 0\}$ such that
\beq{resonance}
	m \cdot \omega = n \;.
\eeq
The \emph{resonance module} of a given $\omega$ is a sub-lattice of $\bZ^d$ defined by 
\beq{module}
	\cL(\omega) \equiv \{ m \in \bZ^d: m\cdot\omega \in \bZ \} \;.
\eeq
The dimension of this module (the number of independent $m$-vectors in $\cL$) is the \emph{rank} of the resonance. An incommensurate or nonresonant rotation vector corresponds to the rank-zero case, $\cL = \{0\}$. 
The \emph{order} of a resonance is the length of the smallest nonzero vector in $\cL$, 
though of course this depends upon the norm used; we will typically use the sup-norm.
The set of resonant frequencies
\[
   \cR \equiv \left\{\omega \in \bR^d: m\cdot \omega =n \mbox{ for some }(m,n)\in \bZ^d \times \bZ \setminus \{0,0\} 
   			\right\}
\]
is the resonance web; it is a dense subset of $\bR^d$. 

For the case at hand, $d=2$ and $k=1$, the image of $\Omega$ in \Eq{integrable} is a curve, as sketched in \Fig{cantorset}. As implied by KAM theory, we expect that, upon a general, smooth perturbation, only tori with ``sufficiently" nonresonant rotation vectors (in the sense of a Diophantine condition) will persist when \Eq{integrable} is perturbed \cite{Cheng90b, Xia92}, and that tori near low-order resonances will be destroyed most quickly. The set of Diophantine rotation vectors is a Cantor set, approximated by the white region in \Fig{cantorset}. The frequencies of the preserved set of tori correspond to a Cantor set \emph{near} $\Omega(z)$: in contrast to the symplectic case, these need not be in the image of $\Omega$.

\InsertFig{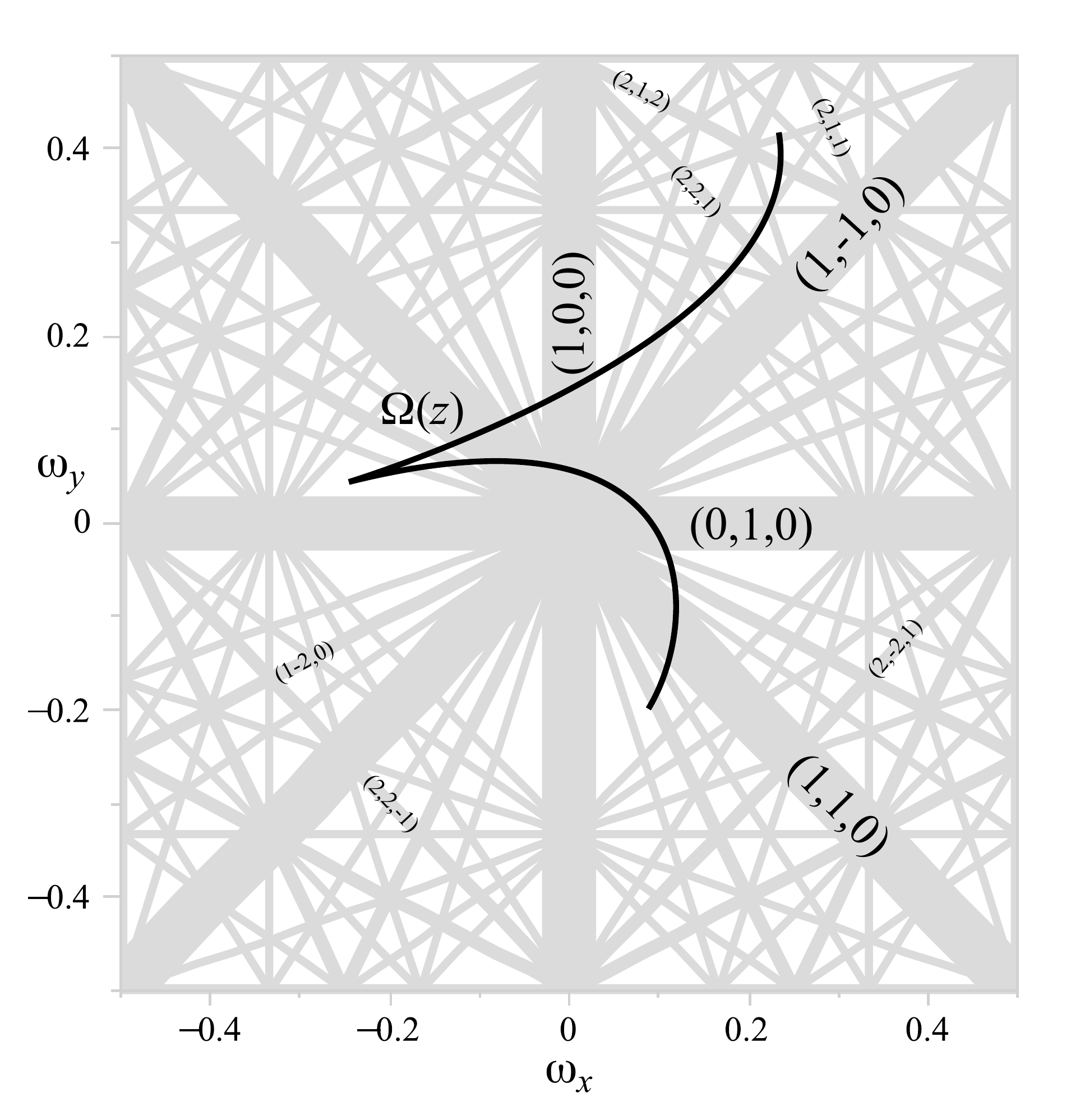}{Sketch of the resonance web, thickened by the Diophantine condition for a one-action map on $\bT^2 \times \bR$. The complement is the positive measure Cantor set of Diophantine rotation vectors. Also shown is the image, $\Omega(z)$, of a frequency map.}{cantorset}{4in}

One way to find invariant tori is to numerically estimate the frequency map; orbits on which the frequency map is well-behaved should correspond to tori or periodic orbits \cite{Laskar93a}. Numerically, we can estimate the rotation vector \Eq{rotationNumber} as
\beq{ApproxFreq}	
	\omega_T(x_0,z_0) = \frac{x_T - x_0}{T}
\eeq
for some $T$. The finite-time estimate \Eq{ApproxFreq} can be used to obtain an approximation of the image of the frequency map on the set of rotational tori.
Since each rotational invariant torus will intersect every angle, we fix can the initial angles $x_0$, and think of \Eq{ApproxFreq} primarily as a function of the initial action
\[
	\omega_T(x_0, \cdot ): \bR^k \to \bR^d.
\]
The estimate \Eq{ApproxFreq} is rather crude since errors will decay as $T^{-1}$ even when the limit exists. There are more sophisticated methods for a single frequency \cite{Efstathiou01,Seara06}, but as these are based on continued fractions that rely on one-dimensional ordering, they cannot be used when $d>1$. Filtered Fourier methods \cite{Laskar93a} are also more accurate; however, these suffer from the difficulty of identifying which peaks correspond to the ``primary" rotation vectors.

\section{Tangency Normal Form}\label{sec:Tangency}

Generically the resonance web will intersect the curve $\Omega(z)$ at a dense set of points. For a rank-one resonance the typical behavior is exemplified by a three-dimensional normal form obtained in \cite{Dullin10a} that is analogous to Chirikov's ``standard map" \cite{Chirikov79} and Froeshl\'e's four-dimensional symplectic map \cite{Froeschle73}. This normal form corresponds to  \Eq{perturbed} with $(x,z) \in \bR^2 \times \bR^1$
and the frequency map and force
\bsplit{Parabola}
	\Omega(z) &= (z+\gamma, -\delta + \beta z^2)\;, \\
	g(x) &= a\sin(2\pi x_1) + b \sin(2 \pi x_2) + c \sin(2\pi(x_1-x_2)) \;.
\end{split}\eeq
Note that $\Omega$ satisfies the nondegeneracy condition \Eq{Wronskian} since the vectors $D\Omega$ and $D^2\Omega$ are never parallel:
\[
	\det(D\Omega, D^2\Omega) = 2\beta \;.
\]
Moreover, since $g \in C^3(\bT^2)$ and has zero average, KAM theory is applicable to \Eq{Parabola}.

The main resonances for \Eq{Parabola} occur when $\Omega$ is commensurate with one of the ``forced" resonances in $g$: the amplitudes $a,b,c$ represent resonant forcing with $m=(1,0)$, $(0,1)$, and $\pm( 1, -1)$, respectively. To limit the number of parameters, we select
\beq{StdParams}
	\gamma = \tfrac12(\sqrt{5}-1) \approx 0.61803 \;, \quad
	\beta = 2 \;, \quad
	a = b = c = 1.0 \;,
\eeq
and vary $\delta$ and $\eps$ for our computations.

\InsertFigTwo{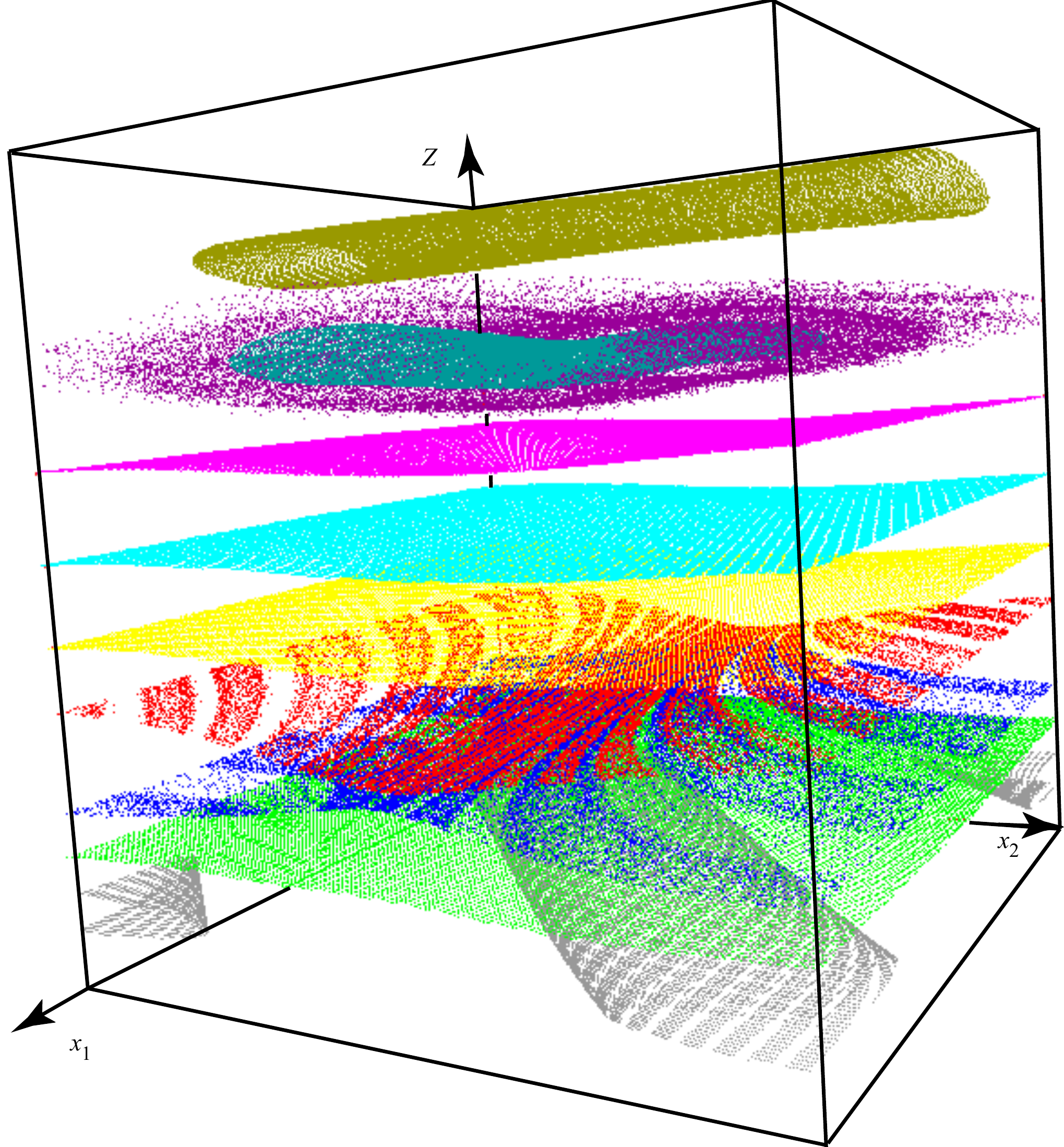}{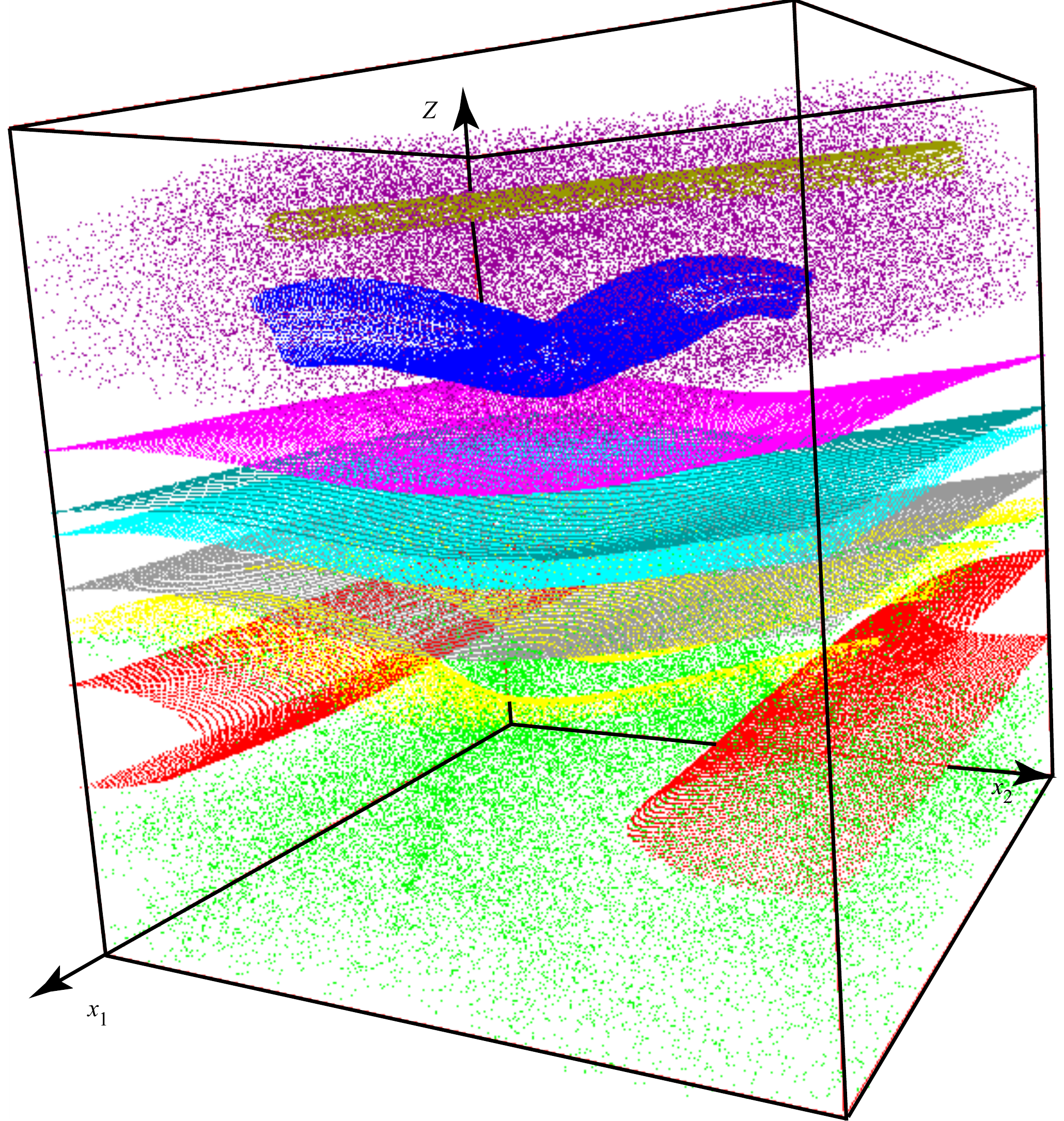}{Some orbits of \Eq{perturbed} with \Eq{Parabola} and \Eq{StdParams}, $\delta = 0.1$ and $\eps = 0.005$ (left), $\eps = 0.015$ (right). The outlined cube is $[-0.5,0.5]^3$. In the left panel, orbits trapped in the forced $(1,0,1)$,  $(0,1,0)^+$, and  $(1,-1,0)^-$ resonances form gold, blue and grey tubes while the red and blue orbits are near the separatrix of the lower $(0,1,0)$ resonance. In the right panel, the $(1,0,1)$ and $(0,1,0)^+$ and the $(0,1,0)^-$ and $(1,-1,0)^-$ resonance pairs overlap to form large chaotic layers.}{Delta01PhaseSpace}{3in}

The locations of the forced resonances for the integrable case, $z^*(m,n)$, are given in \Tbl{resonances}. Note that most resonances occur at two $z$-locations for each $\delta$ because $\Omega$ is parabolic. If $\Omega$ is tangent to a resonance line, there will be only one $z$-location, and the structure of the dynamics near this point resembles a nontwist area-preserving map \cite{Dullin10a}. Under the approximation that only one resonance is present and its amplitude is small, its width $w$---defined to be the maximum difference between the upper and lower separatrices---can be computed by standard techniques \cite{Dullin10a}. As an example, the phase portraits for $\delta = 0.1$ in \Fig{Delta01PhaseSpace} show four prominent resonances; the locations and widths of these forced resonances are given in \Tbl{locations}. When $\eps$ is small, as in the left pane, much of phase space away from these resonances is foliated by rotational tori; the region within the width of each forced resonance is predominantly foliated by ``librational" tori, the tubes in the figure. Of course, the dynamics near the resonance separatrices is chaotic, and as $\eps$ increases, chaotic orbits become more prominent. 

\begin{table}[tbph]
   \centering
   \begin{tabular}{@{} cccc @{}} 
      \toprule 
       $(m,n)$   & $z^*$ & $w^2$ & $\delta_T$ \\
      \midrule
      $(1,0,n)$	& $n-\gamma$ &  ${\frac{8\eps a}{\pi}}$ & none \\ 
      $(0,1,n)^\pm$	& $\pm \sqrt{\frac{\delta+n}{\beta}}$ 
      			    & ${\frac{ 4\eps b}{\pi\beta z^*}}$
				    & $-n$ \\
      $(1,-1,n)^\pm$	& $\frac{1}{2\beta}
      				\left[1 \pm\sqrt{1+4\beta(\delta+\gamma-n)}\right]$
					& ${\frac{8\eps c}{\pi(1-2\beta z^*)}}$
					& $n -\gamma-\frac{1}{4\beta}$ 
					\\ 
      \bottomrule
   \end{tabular}
   \caption{Locations of the forced resonances $m \cdot \Omega(z^*) = n$ for the frequency map \Eq{Parabola}, their full widths $w$, and values of $\delta$ at which $\Omega(z)$ is tangent to the resonance.}
   \label{tbl:resonances} 
\end{table}

\InsertFigTwo{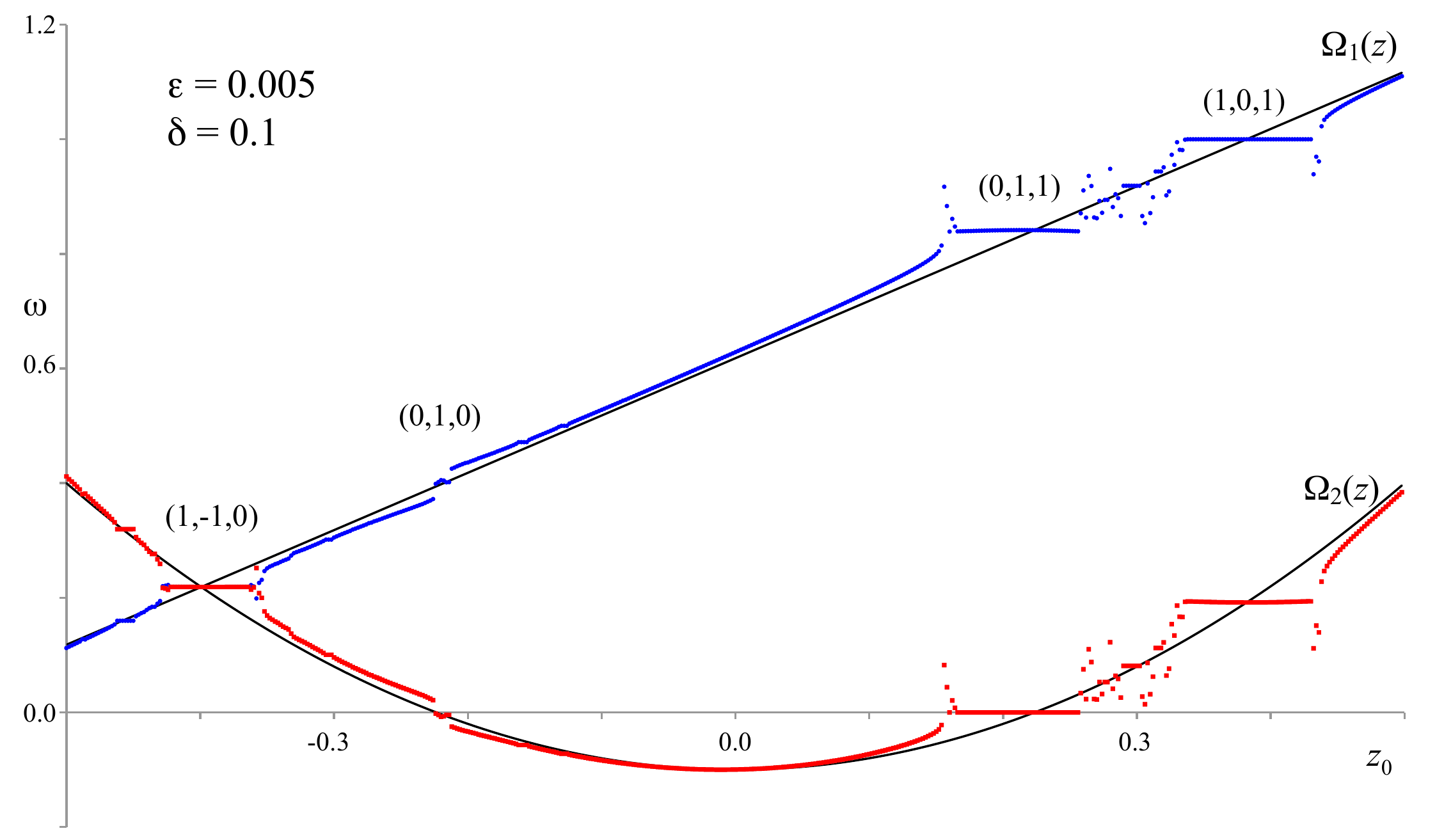}{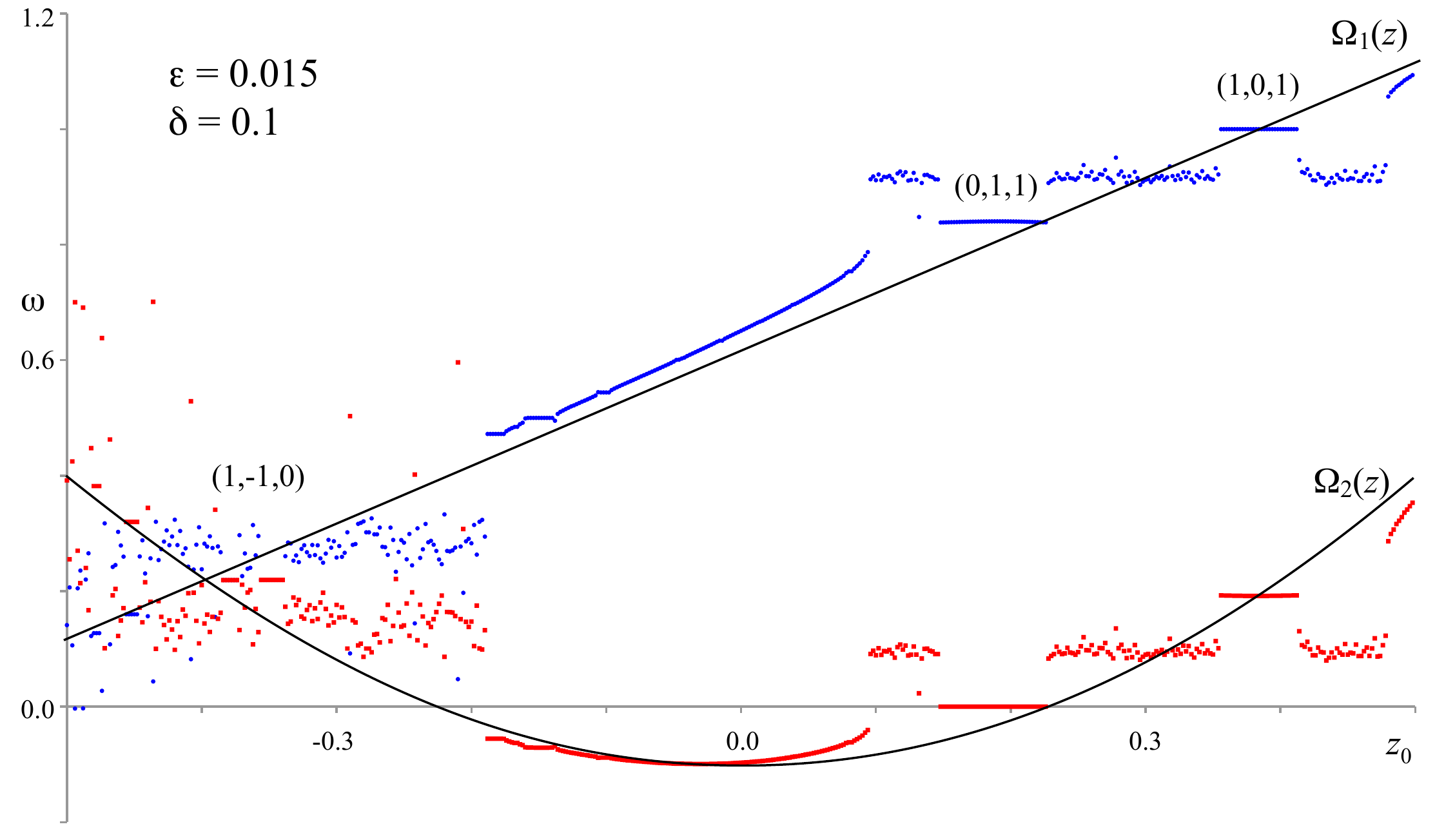}{Numerically computed frequencies, $\omega_T$, with $T=10^5$. Parameters are the same as \Fig{Delta01PhaseSpace}. The initial conditions are $(0,0,z_0(j))$ with with $z_0(j) = -0.5+ \frac{j}{500}$, $j = 0,1, \ldots,500$. The dots are the first (blue) and second (red) components of $\omega_T$ as a function of $z_0$, and the curves are the components of $\Omega(z)$ of \Eq{Parabola}.}{FreqMapVsZ}{3in}

A numerical computation of the approximate, perturbed frequency map using \Eq{ApproxFreq} for $T=10^5$ is shown in \Fig{FreqMapVsZ} for the same parameters as \Fig{Delta01PhaseSpace} and with initial conditions along the line segment $\{(0,0,z_0): z_0 \in[-0.5,0.5]\}$. When $\eps = 0.005$, as in the left pane, the components of $\omega_T$ follow those of $\Omega(z)$ closely except near the forced resonances where the frequencies resonantly lock within a width $w$ about $z^*$. Locking intervals are visible for the $(1,0,1)$, $(0,1,0)^+$ and $(1,-1,0)^-$ resonances; the fourth resonance $(0,1,0)^-$ appears as a jump in $\omega_T$ instead of a locking interval because the chosen line of initial conditions passes near the hyperbolic invariant circle for that resonance, as can be seen in \Fig{Delta01PhaseSpace}. Forced resonances in \Tbl{locations} with other values of $n$ do not occur within the domain of the figure.

Though the frequency components within a mode-locking interval appear constant on the scale of \Fig{FreqMapVsZ}, they actually vary slowly, maintaining the resonance condition across the interval.

\begin{table}[tbph]
   \centering
   \begin{tabular}{@{} l|cc|cc @{}} 
      \toprule 
       $(m_1,n_1)$  & $z^*_{theory}$ & $z^*_{expt}$ & $w_{theory}$ &  $w_{expt}$ \\
      \midrule
      $(1,0,1)$	&   0.382 &0.351 &0.113 & 0.094\\ 
      $(0,1,0)^+$&  0.224 & 0.202 & 0.169 & 0.112\\
      $(0,1,0)^-$& -0.224 & -0.224 & 0.169 & \\
      $(1,-1,0)^-$& -0.399 & -0.384 & 0.084 &0.058\\ 
      \bottomrule
   \end{tabular}
   \caption{Comparison of the theoretical and experimental locations (estimated as the centers of the flat regions) and widths of the forced resonances for $\delta =0.1$ and $\eps = 0.005$ as in the left pane of \Figs{Delta01PhaseSpace}{FreqMapVsZ}. The width of the $(0,1,0)^-$ resonance cannot be measured along this curve of initial conditions}
   \label{tbl:locations} 
\end{table}

Chaos is much more visible in the right panes of Figs.~\ref{fig:Delta01PhaseSpace}--\ref{fig:FreqMapVsZ} where $\eps = 0.015$; in particular, the frequencies vary irregularly with $z_0$ for $z_0 < -0.292$, indicative of the strong chaos associated with overlap of the  $(0,1,0)^{-}$ and $(1,-1,0)^-$ resonances. However, for $z_0 \in [-0.29,0.092]$ both components of $\omega_T$ vary relatively smoothly; the corresponding phase portrait shows many tori in this region. The values of each component in \Fig{FreqMapVsZ}(b) are visibly shifted to the left from the unperturbed frequency \Eq{Parabola}; indeed the extremum occurs at $z_0 = -0.03$ so that $\omega_T(0,0,-0.03) \approx \Omega(0)$. For $z_0 > 0.092$, the frequencies alternately exhibit resonant trapping and chaotic layers.

\InsertFig{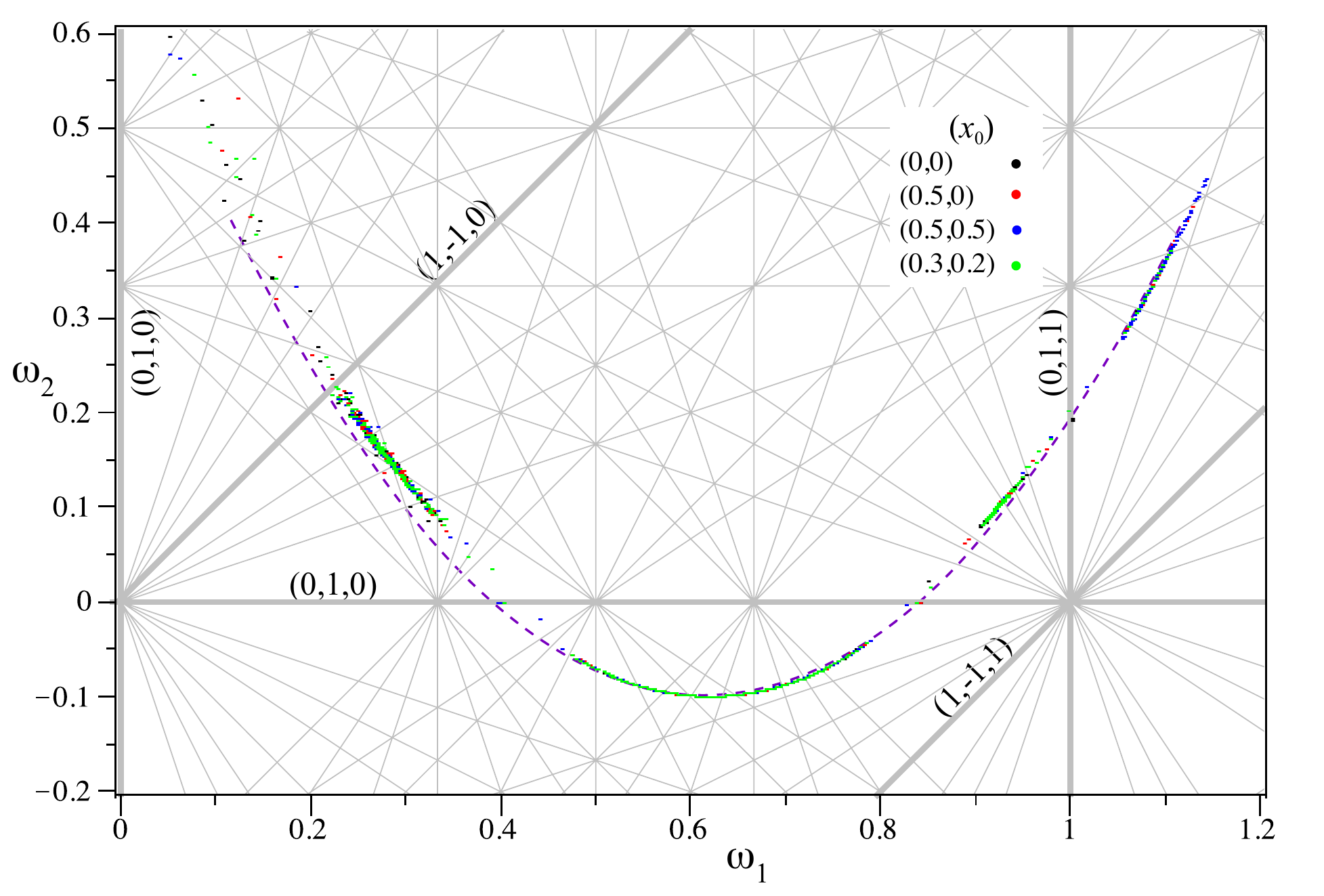}{Numerically computed frequency map for \Eq{Parabola} with \Eq{StdParams}, $\eps = 0.015$ and $\delta = 0.1$, as in the right pane of \Fig{FreqMapVsZ}. The four sets of points correspond to initial conditions on lines $(x_0,z_0)$ with $(x_0)$ fixed as indicated and $z_0$ ranging from $-0.5$ to $0.5$ in steps of $0.02$. The dashed curve (purple) represents the image of unperturbed frequency map \Eq{Parabola} for the same range of $z$. Resonance lines up to order $3$ are shown in grey; the forced resonances are the thicker lines.}{FreqMapVaryx0}{5in}

The chaos visible in \Fig{FreqMapVsZ}(b) is less obvious in the image of the approximate frequency map $z \mapsto \omega_T$ shown in \Fig{FreqMapVaryx0}. The computed frequencies are close to the unperturbed parabolic curve \Eq{Parabola}; however, they fall visibly above $\Omega(z)$ in the chaotic region near the overlapping $(1,-1,0)^-$ and $(0,1,0)^-$ resonances and they extend considerably beyond the left endpoint of the unperturbed curve, $\Omega(-0.5)$.

Computations of $\omega_T$ for four different initial angles are shown in \Fig{FreqMapVaryx0}. In most cases the four computed frequency curves are nearly identical even though a trajectory with a given $z_0$ may have a significantly different computed frequency for each initial angle. Indeed, in the regular regions the computed frequencies for the same $z_0$  but different angles often differ by $10\%$ even though they typically lie within $1\%$ of the parabola \Eq{Parabola}. This is precisely as expected for a region foliated by rotational tori intersecting different angles at different heights. The difference between the initial angles is especially prominent near resonances. For example, the initial phases $(0.5, 0.5)$  and $(0.3,0.2)$ cross the tube of librating orbits for the $(0,1,0)^-$ resonance and they have mode-locked segments with $\omega_T \approx(\frac25,0)$; however the phases $(0,0)$ and $(0.5,0.5)$ appear to lie near the hyperbolic circle of this resonance and their corresponding frequencies jump as $z_0$ moves across the resonance.

Variation of the image of the frequency map with $\eps$ is shown in \Fig{FreqMapDelta0}  for $\delta = 0.0$ and initial conditions on the line $(0,0,z_0)$. 
For this value the parabola $\Omega(z)$ is tangent to the $(0,1,0)$ resonance. 
When $\eps = 0.001$, the numerical frequency map (the black dots) is indistinguishable from the parabola $\Omega(z)$, except for some small deviations when $z_0 \in [-0.372, -0.344]$ where the orbits appear to be trapped near the $(1,-1,0)^-$ resonance with $\omega_T \approx (\frac{8}{31},\frac{8}{31})$, when $z_0 \in [0, 0.08]$ where $\Omega(z)$ is near the $(0,1,0)$ resonance, and when $z_0 \in [0.358,0.404]$ where it is near the $(1,0,1)$ resonance. 
Other resonance lines are shown in the figure, but when the perturbation is this small, they have little effect on the image, indicating that most of phase space is foliated by invariant tori. For $\eps = 0.01$ much of the image still follows the parabola, but for $\eps \ge 0.02$ the image becomes increasingly irregular, correlating with regions of chaotic behavior in the phase space. Indeed, when $\eps = 0.02$ there appear to be no rotational invariant tori except for  $z\in [0.091, 0.153]$. When $\eps = 0.03$, 
there appear to be no rotational invariant tori for  $z_0 \in [-0.5,0.5]$. We will explore this transition in the next section.

\InsertFig{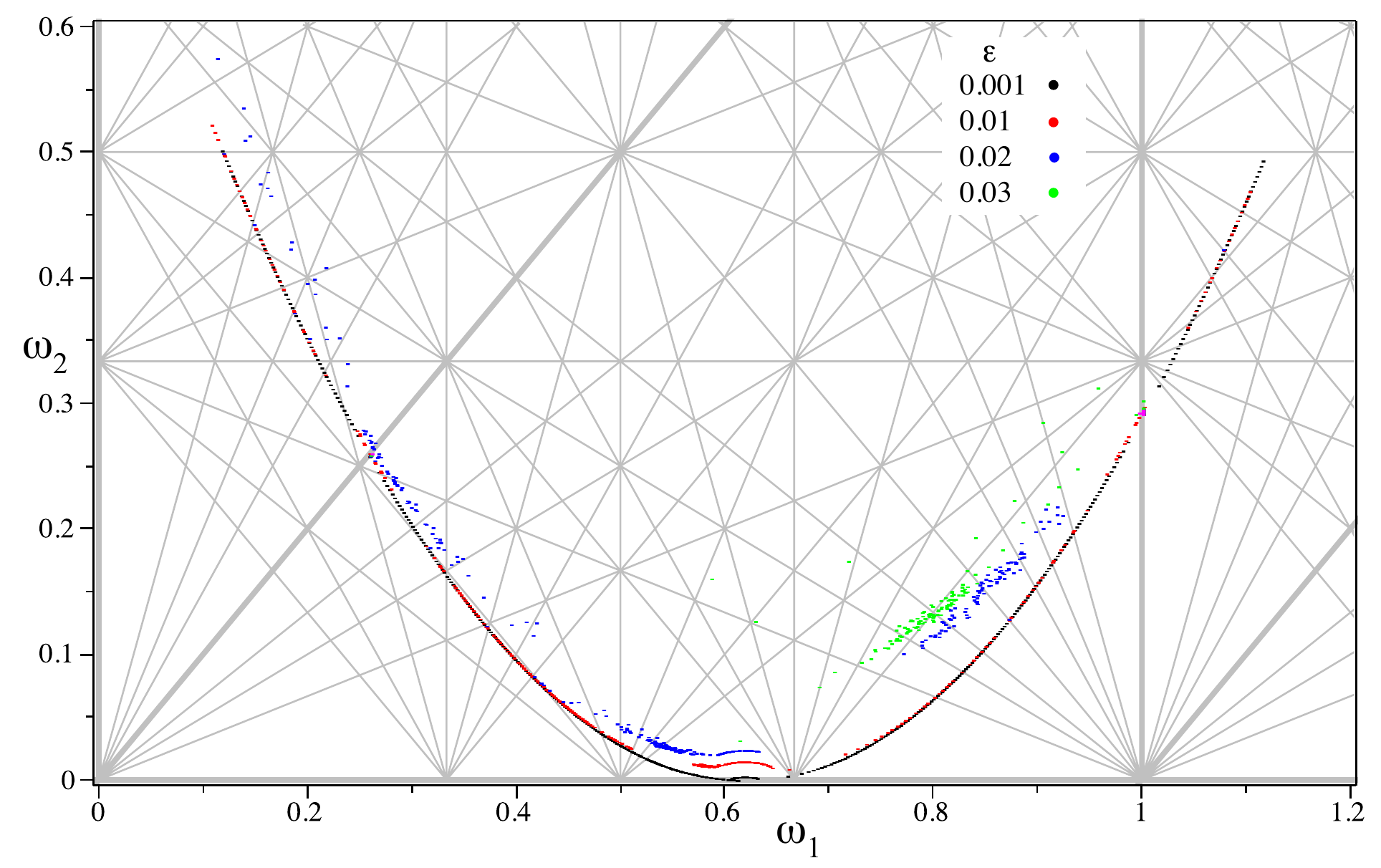}{Approximate images of the frequency map \Eq{ApproxFreq} for $T=10^5$ iterates for \Eq{Parabola} with \Eq{StdParams}, $\delta = 0$ and $a=b=c=\eps$. The value of $\eps$ ranges from $0.001$ to $0.03$ as indicated.}{FreqMapDelta0}{5in}

Finally, the variation of the frequency map with $\delta$ is shown \Fig{FreqMapVaryDelta}. In this figure, gaps around the forced resonances are prominent, and smaller gaps around some of the higher order resonant lines, e.g. $(1,2,0)$ and $(2,1,0)$ can also be seen.
 
\InsertFig{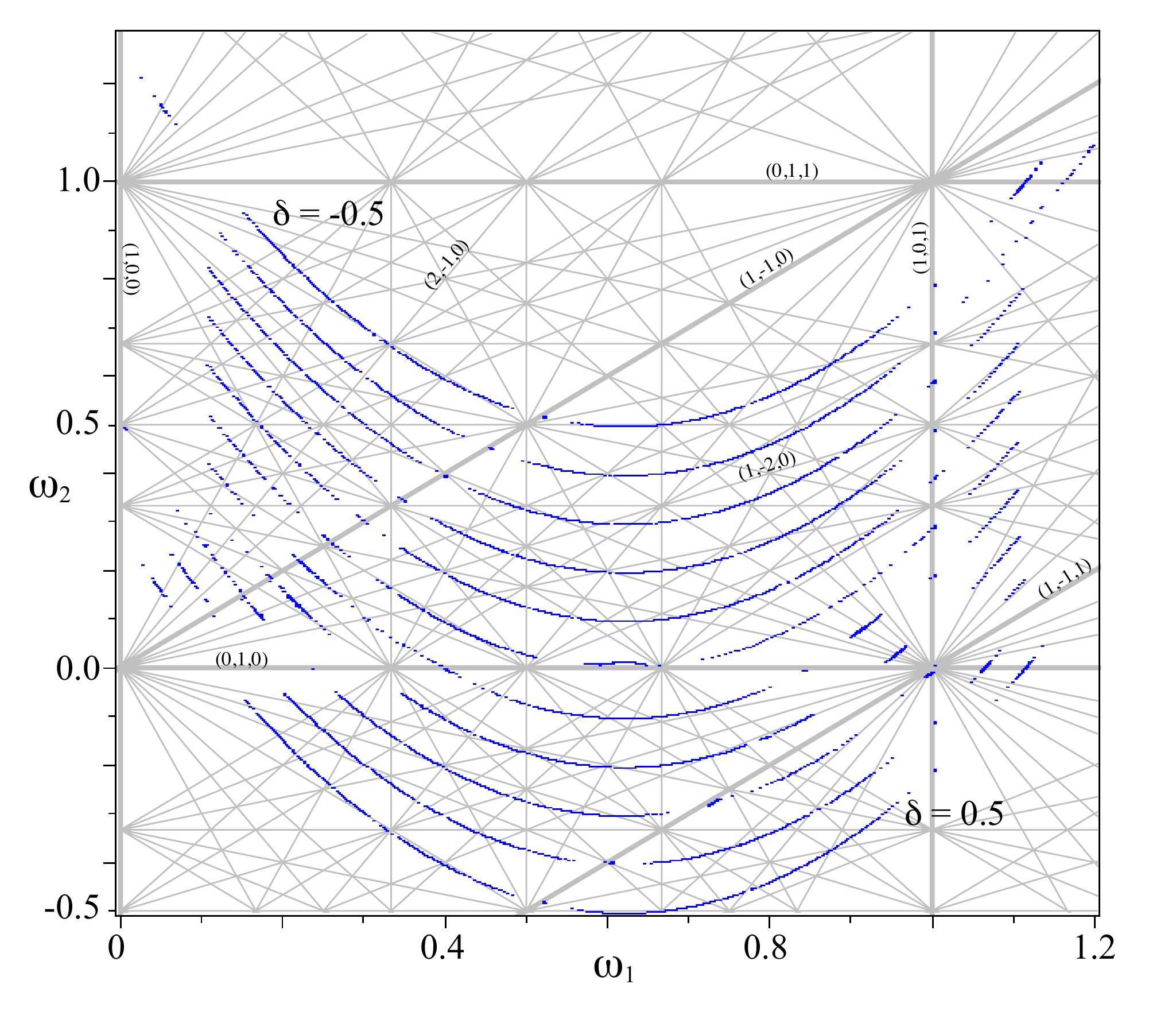}{Images of the approximate frequency map for \Eq{Parabola} with \Eq{StdParams}, $\eps = 0.01$, $T=10^5$, and varying $\delta$ from $-0.5$ to $0.5$ in steps of $0.1$. The initial angle is $(0,0)$, and $z_0$ varies from $-0.5$ to $0.5$. Note that the curve for $\delta = -0.5$ is equivalent to that for $\delta = 0.5$ under $\omega_2 \to \omega_2+1$}{FreqMapVaryDelta}{5in}

\section{The Last Invariant Torus}\label{sec:lastTorus}

In 1968 John Greene developed a method for determining the ``last" rotational invariant circle for an area-preserving map. Greene demonstrated that a circle with a given rotation number $\omega$ exists when a sequence of periodic orbits whose rotation numbers limit on $\omega$ have bounded ``residue". 
He provided compelling evidence that the last circle for Chirikov's standard map has the golden mean rotation number, that the ``critical" invariant circle is not $C^2$, and the phase space in its neighborhood exhibits self-similarity \cite{Greene79,Greene80}. 
This self-similarity leads to a renormalization group explanation for the existence of invariant circles \cite{MacKay93,Arioli10}.  
More generally, it is hypothesized that the most robust invariant circles for twist maps have ``noble" rotation numbers \cite{Schmidt82,Escande84, Buric90, MacKay92b}. 
Though there have been many attempts to generalize Greene's ideas to higher dimensions---in particular for symplectic maps \cite{Bollt93, Vrahatis97, Kurosaki97, Haro99, Tompaidis96, Tompaidis99, Chandre00, Chandre01, Zhou01b,Celletti04}---there is still no comparable theory that describes the robustness and destruction of multi-dimensional tori.

The question of which torus is the ``last" is perhaps of less importance for symplectic maps because when $d=k>2$, the tori do not form barriers to transport. When the tori have codimension-one, however, the question is more relevant.

As $\eps$ increases in \Eq{perturbed}, rotational tori become less common.
It appears that there is an $\eps_c(\beta,\delta,\gamma)$, such that whenever $\eps > \eps_c$ there are no rotational invariant tori.  Which is the ``last" torus and what are its properties?

To make this question easier to investigate, it helps to add a vertical translation symmetry to the map by assuming that there is a $j \in \bZ^d$ such that
\beq{Dehn}
	\Omega(z+1) = \Omega(z) +j \;.
\eeq
In this case, \Eq{perturbed} is isotopic to a three-dimensional version of the \emph{Dehn twist}. For each orbit $\{(x_t,z_t): t \in \bZ\}$ of the lift of such a map, the set $\{ (x_t+kjt,z_t+k): t\in \bZ\}$ is also an orbit for any $k \in \bZ$; consequently, the phase space structure of the map on $\bT^d \times \bR$ is periodic in $z$.

For our numerical experiments we set
\beq{periodicOmega}
	\Omega(z) = (z +\gamma, -\delta +\lambda\sin^2(\pi z)) \;,
\eeq
which has a parabolic form like that of \Eq{Parabola} when $|z| \ll 1$. This map satisfies \Eq{Dehn} with $j = (1,0)$. Since the most robust tori of \Eq{Parabola} appear to occur near $z = 0$, the last torus for \Eq{periodicOmega} should be close to that for \Eq{Parabola} for $\beta = \lambda \pi^2$. We choose $g$ from \Eq{Parabola}, as before and set
\beq{StdParamsII}
	\lambda = 1,\; \gamma = \tfrac12(\sqrt{5}-1), \; a=b=c=1.0\;,
\eeq
so that $\Omega_2$ varies over a unit interval. Periodicity in $x$ implies that $\gamma$ and $\delta$ can be restricted to any unit interval; thus we will study $\delta \in [-0.5,0.5]$.

\InsertFigTwo{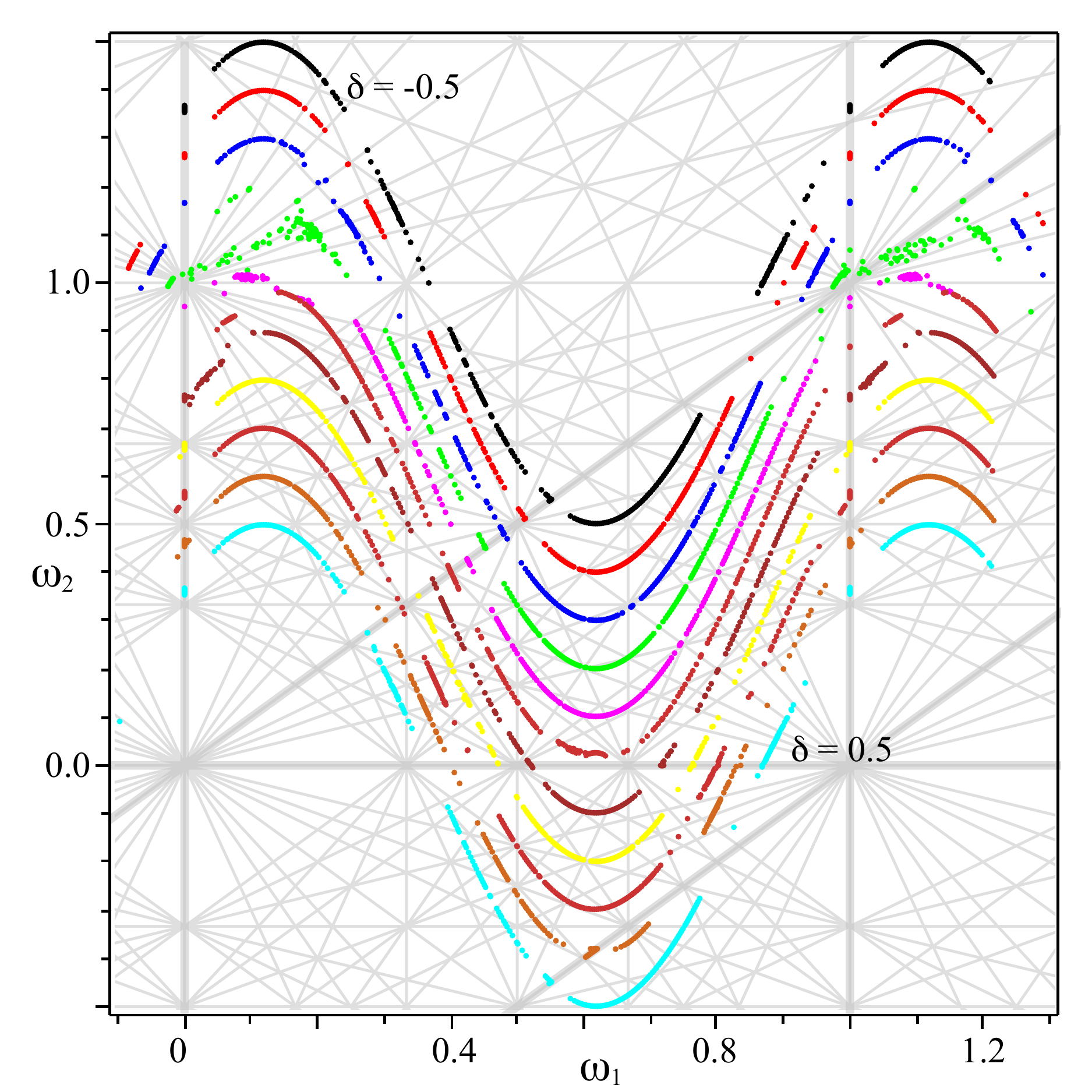}{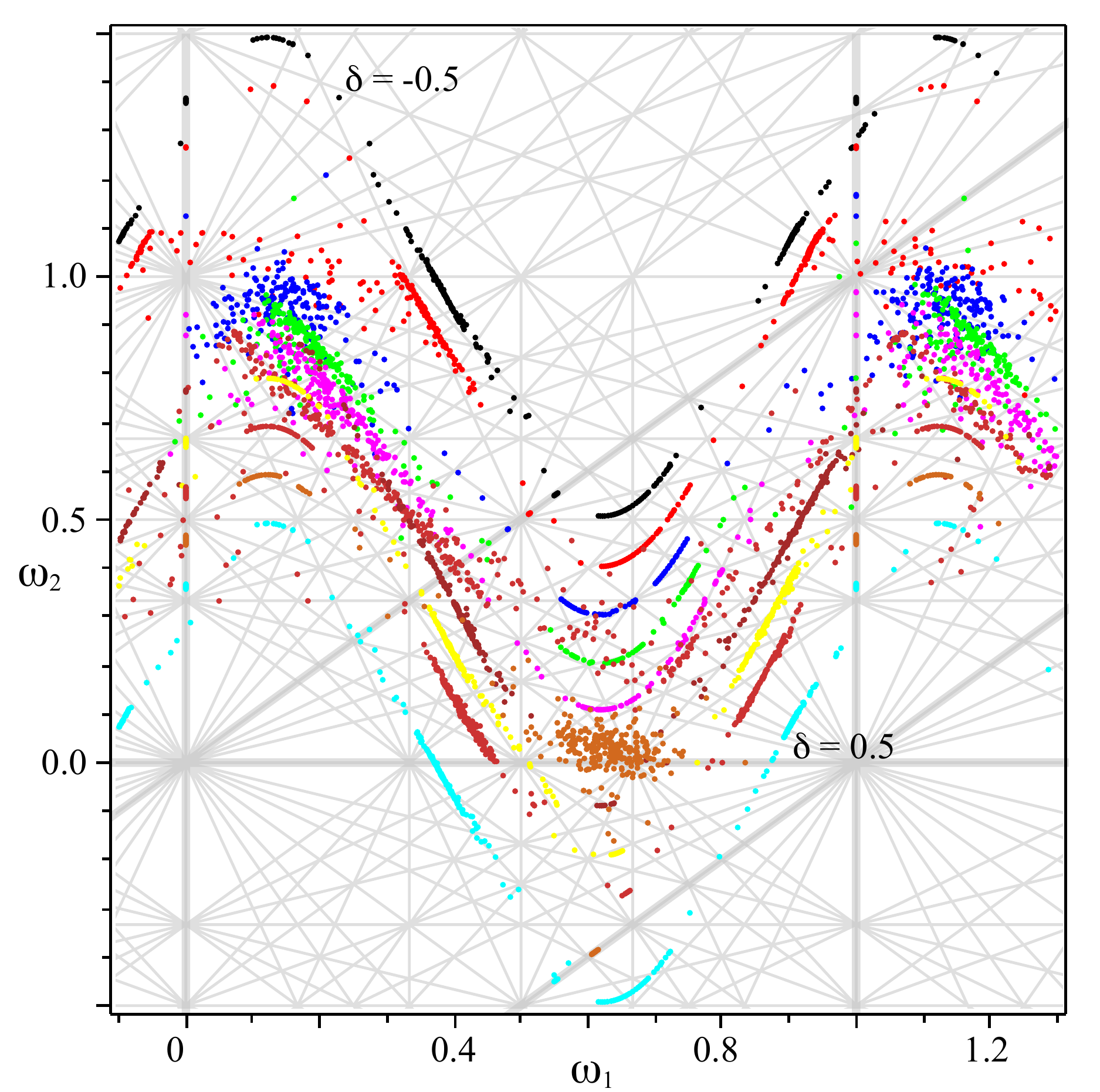}{Frequency Maps for \Eq{perturbed} with \Eq{periodicOmega}, with  \Eq{StdParamsII} with $T = 10^5$. The curves correspond to $\delta$ ranging from $-0.5$ (black) to $0.5$ (cyan) in steps of $0.1$. For the left pane, $\eps = 0.01$, and for the right, $\eps = 0.026$.}{PeriodicFreqMap}{3in}

Frequency maps for \Eq{perturbed} with \Eq{periodicOmega} and the force $g$ in \Eq{Parabola} at two values of $\eps$ are shown in \Fig{PeriodicFreqMap}. Since the map for $\eps = 0.01$ (left pane) is highly regular it indicates there are many invariant tori; however the irregularity of the frequencies for $\eps = 0.026$ (right pane), indicates that most tori have been destroyed.

\subsection{Crossing Time}\label{sec:crossingTime}
The simplest rigorous criterion that forbids the existence of invariant tori is the existence of ``climbing orbits"; i.e., there are no rotational tori whenever there is an unbounded orbit. Since it is numerically impossible to verify that an orbit is unbounded, an implementation of this idea requires a bound on the vertical extent of rotational tori. Define the \emph{vertical extent} of a set $\cT \subset \bT^d \times \bR$ as
\[
	\Delta(\cT) = \max_{(x,z) \in \cT}(z) - \min_{(x,z) \in \cT}(z) \;.
\]
Rotational tori can be shown to have a bounded vertical extent for the case of 
symplectic maps with uniformly nondegenerate twist; indeed, ``Birkhoff's Second Theorem" implies that a Lagrangian, rotational invariant torus on which the dynamics is minimal must be a Lipschitz graph over the angle variables \cite{Bialy92, Bialy04, Arnaud10}. The vertical extent of any such torus is obtained from the Lipschitz constant. This theorem justifies ``converse-KAM" theory and can be used both analytically and numerically to rule out the existence of rotational tori \cite{MacKay85,MacKay89}. Unfortunately, there can be no straightforward generalization of Birkhoff's theorem to a volume-preserving map of the form \Eq{perturbed} since its rotational invariant tori are not always graphs  \cite{Dullin10a}. Such ``meandering" tori (like those of nontwist area-preserving maps \cite{Simo98, Wurm05}) often occur near resonant tangencies of the frequency map.

Even though we  know of no rigorous bound on the vertical extent of tori of \Eq{perturbed}, we will assume that one does exist, i.e., there is a $\Delta_{max}$ such that $\Delta(\cT) < \Delta_{max}$ for all rotational invariant tori $\cT$. In this case, using the vertical periodicity of \Eq{periodicOmega}, rotational tori can be ruled out if there is an orbit that climbs a distance $1+\Delta_{max}$. Numerical observations indicate that $\Delta_{max}$ is quite small. The largest observed vertical extent, $\Delta(\cT) \approx 0.2$, occurs near a reconnection bifurcation \cite{Dullin10a}. 

Let the \emph{crossing time}, $t_{cr}$, of an initial condition be the minimal time over which $z$ changes by $1+\Delta_{max}$:
\[
	t_{cr} = \min \{t>0: |z_{t}-z_0| \ge 1+\Delta_{max} \} \;.
\]
We compute the crossing times with $\Delta_{max} =0.3$ for a set of $k=10$ initial conditions of the form $(0,0,z_0)$ obtained by varying $z_0$ in a narrow range about $-\gamma$ where there is a hyperbolic invariant circle (for $\eps  \ll 1$) due to the $(1,0,0)$ resonance.
To make the computation finite, the number of iterations for each initial condition was limited to at most $10^{10}$.

We observed that the crossing time grows rapidly as $\eps$ decreases and appears to go to infinity at some critical value, $\eps_c$, with a power-law singularity of the form
\beq{polefit}
	t_{cr} \sim T(\eps;K,\eps_c,\alpha) \equiv \frac{K}{(\eps-\eps_c)^{\alpha}} \;.
\eeq
An example is shown in \Fig{CrossingTime}. If this form is correct, then $\eps_c$ is the parameter at which the last invariant torus is apparently destroyed.

\InsertFig{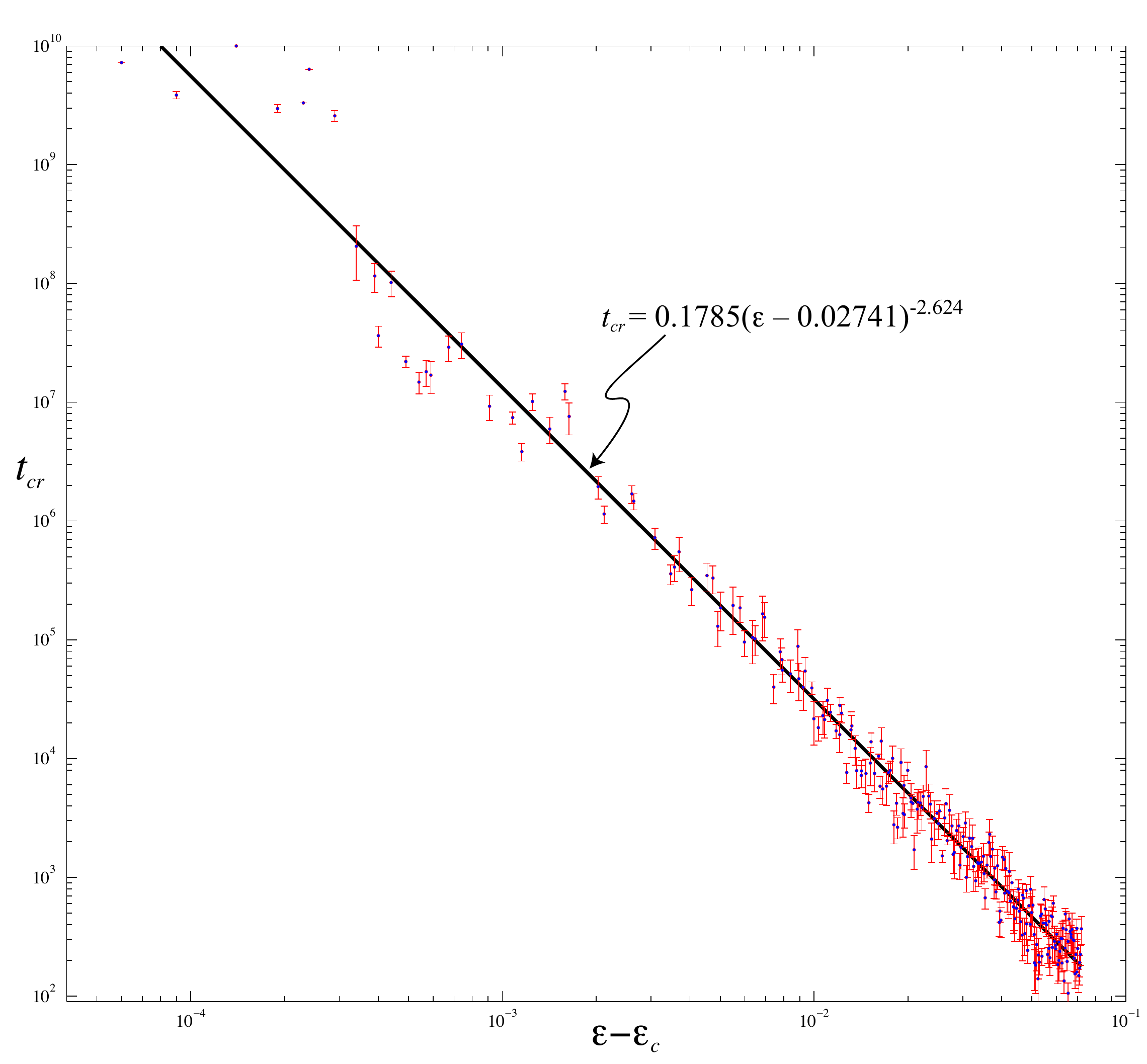}{Log-Log plot of the median crossing time (points) for $\delta = 0.1$ and $\Delta_{max} =0.3$ as a function of $\eps-\eps_c$ where $\eps_c = 0.02741$. The error bars show the median absolute deviation, $t_{mad}$, for $10$ trials at each $\eps$ value. The line is an unbiased fit to the logarithm of \Eq{polefit} that minimizes the log-mean square error \Eq{meanSquareError}.}{CrossingTime}{4in}

When $\delta = 0.1$ and $\eps \ge 0.0273 = \eps_U$, there were orbits in the trial set with $t_{cr} < 10^{10}$, thus $\eps_c < \eps_U$. However, when $\eps \in[0.02733,0.0277)$, $t_{cr} > 10^{10}$ for some orbits in the trial set; consequently, our computations cannot estimate the average crossing time in this range. Since the median can still be computed if the majority of trial orbits cross, we used the median crossing time to fit the data using  $k=10$ trials at each of $N=238$ values of $\eps \in [0.02747,0.1]$.  The spread in the data is estimated using the \emph{median absolute deviation} 
\[
	t_{mad} = \med_{j=1\ldots k}\{|t_j-\med_{i=1\ldots k}(t_i)|\} \;.
\]

The three parameters, $K$, $\eps_c$, and $\alpha$ of \Eq{polefit} were fit by  minimizing the squared log-error:
\beq{meanSquareError}
	E^2 = \frac{1}{N}\sum_{i}\left[\log t_{cr}(\eps_i)-\log T(\eps_i;K,\eps_c,\alpha) \right]^2 \;.
\eeq
The resulting fit, shown in \Fig{CrossingTime} for $\delta = 0.1$, gives $\eps_c = 0.02741$, and an exponent $\alpha = 2.624$ with an rms error $E = 0.22$. However, this estimated value of $\eps_c$ is clearly too large since we observed crossing orbits at this parameter value (see also \Sec{exploration}). A two-parameter fit, setting $\eps_c = 0.0273$, a more reasonable value,  gives $\alpha = 2.772$ and $E = 0.24$. 
The average crossing time also appears to fit the model \Eq{polefit}, giving $\eps_c=0.0274$, and $\alpha = 2.617$ for the full fit, and $\alpha = 2.753$ for the fixed-$\eps_c$ fit with similar rms errors.

Though the data provide clear evidence for \Eq{polefit}, it is difficult to compute the exponent accurately. In particular, some of the largest deviations from the fit occur near the singularity: it is possible that if computations could be done for $t_{cr} > 10^{10}$, the exponent would be different. Nevertheless, it is intriguing that the observed exponent for the three-dimensional map is close to that for a critical noble invariant circle of a twist map ($\alpha = 3.05$ \cite{MacKay84}).

Fits to the crossing time data for other values of $\delta$ are similar and in each case there seems to be a power law singularity in the crossing time; however, the exponent varies considerably with $\delta$.

\subsection{Graphical Exploration}\label{sec:exploration}
An alternative technique to estimate $\eps_c$ that also yields bounds on the location of  rotational tori is to determine if orbits lie in a large, connected chaotic zone, or are trapped in a set with small vertical extent and thus presumably bounded by rotational tori.
This method, primarily visual, uses the openGL-based iteration code to display the evolution of an orbit, recall \Fig{Delta01PhaseSpace}. This interactive program displays a trajectory as a continuously evolving ``cloud" of its most recent  $2^{13}$ iterates.

\InsertFig{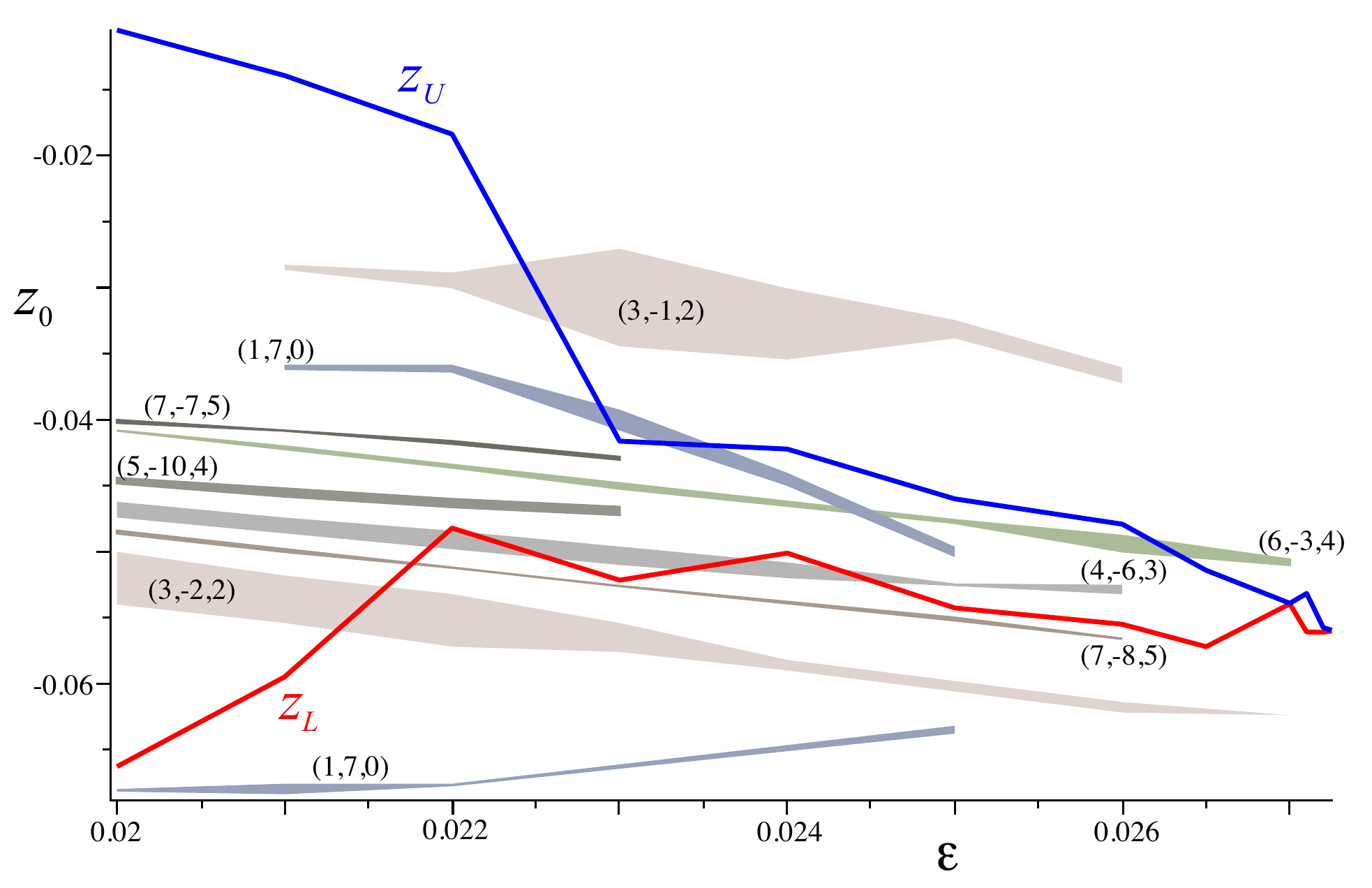}{The interval of rotational tori as a function of $\eps$ for $\delta = 0.1$ on the line $(0,0,z_0)$. There are no rotational tori for $z_0 \not\in (z_L,z_U)$ for initial conditions $(0,0,z_0)$. The vertical extent through $x_0 = 0$ of some low-order resonances, labeled by $(m_1,m_2,n)$,  are also shown. The two $(1,7,0)$ resonances appear to reconnect near $\eps = 0.0255$; unfortunately, their librational tori are destroyed before this happens and so the reconnection is hard to observe.}{TorusBounds}{5.5in}

Using this method, it is quite easy to discover that the region containing rotational tori shrinks to a narrow range of initial conditions as $\eps$ grows. For initial conditions on the line $(0,0,z_0)$ and $\eps = 0.02$, this range is $z_0 \in (z_L, z_U) =  (-0.0663, -0.0104) \pm 0.0001$, as shown in \Fig{TorusBounds}. The experiments indicate that there are tori near the upper and lower bounds of this range. Orbits with $z_0 \in (z_L, z_U)$ may be chaotic or lie on resonant, nonrotational tori; nevertheless, each such orbit appears to be trapped in $(z_L,z_U)$. Moreover, the orbit of $(0,0,z_L)$ is chaotic and was observed to move below $z = -0.5$, while the orbit of $(0,0,z_U)$ is chaotic and moves above $z = 0.5$.

As $\eps$ grows, the interval $(z_L,z_U)$ shrinks as shown in \Fig{TorusBounds}. 
From this data, we assert that there are no invariant tori for $\eps = 0.0273$ because we found a pair of orbits that ``cross" any domain where tori could exist. In particular, the orbit of $(0,0,-0.056002)$ is chaotic and has $\min(z_t) < -0.5$, while the orbit beginning at $(0,0,-0.05612)$ has $\max(z_t) > 0.5$.

Note, however, that the destruction of tori does not appear to be monotonic with $\eps$: in the figure the domain of tori shrinks almost to a point, $(-0.053956,-0.053906)$, at $\eps = 0.0270$, but then grows to $(-0.056080, -0.053160)$ at $\eps = 0.0271$ before finally disappearing at $\eps = 0.0273$. Our observations indicate that there are no rotational tori for larger values of $\eps$, though we cannot rule out tori that might reappear for very small intervals of $\eps$.

These experiments indicate again that $\eps_c \le 0.0273$ for $\delta = 0.1$. For $\eps=0.02729$, the orbit of $(0,0,-0.0561)$, though weakly chaotic, appears to be trapped between rotational tori. This orbit provides an estimate of the rotation vector of the ``last rotational torus", from \Eq{ApproxFreq},
\[
	\omega \approx (0.619,-0.0859)
\]
It is difficult to compute this value more accurately without a more systematic method to locate tori; we plan to do this in a future paper \cite{Fox11}.

Similar experiments for other values of $\delta$ show that the critical $\eps$ value ranges from about $0.0187$ to $0.0347$, with the largest observed value for $\delta = -0.4$.

\InsertFig{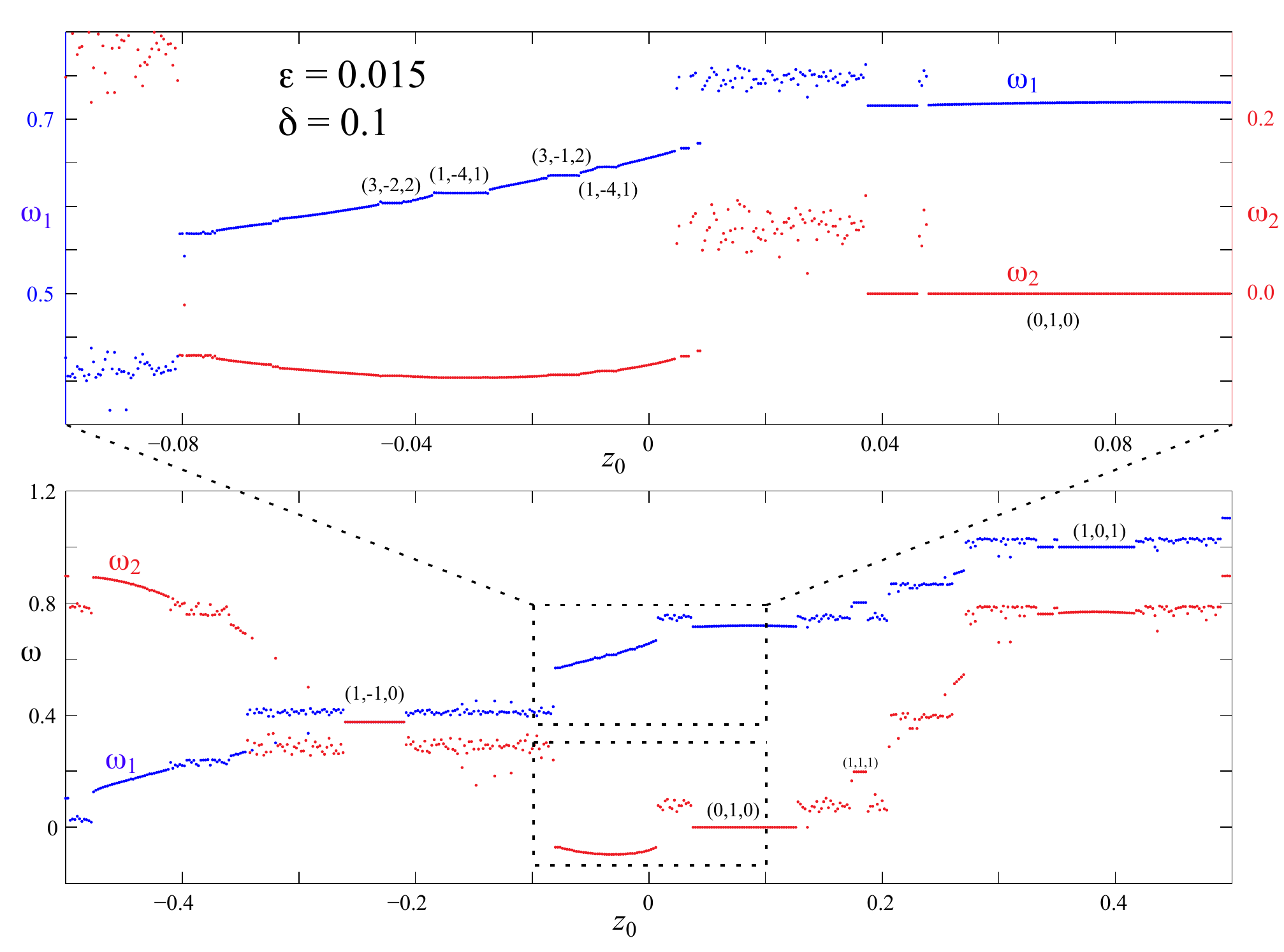}{Numerically computed frequency components \Eq{ApproxFreq} as a function of $z_0$ for $\eps = 0.015$ and $\delta = 0.1$ with $x_0 = (0,0)$ and $T=10^7$ (lower pane) and $T=10^8$ (upper pane) for the frequency map \Eq{periodicOmega}. The relatively smooth variation of the frequencies near $z_0 = -0.45$ and $-0.05$ hints at the existence of rotational tori. The upper pane shows an enlargement of the dashed boxes around $z_0=0$. In the upper pane, the frequency components $\omega_1$ and $\omega_2$ are plotted on different scales as shown on the left and right edges of the graph.}{FreqMapZoomeps015}{6in}

\subsection{Frequency Maps}\label{sec:freqMaps}

As a final method to look for the last torus, we explore in more detail the numerical frequency maps shown in \Fig{PeriodicFreqMap}. Again, we focus on the case $\delta = 0.1$.

It is easier to detect the irregularity of the frequency map from its components; these are shown for $\eps = 0.015$ in \Fig{FreqMapZoomeps015} as a function of the initial $z$-value. Chaotic regions correspond to a scatter in $\omega_T$ values for nearby initial conditions. These are interspersed with intervals where $\omega_T$ appears to be more or less smooth indicating rotational or librational tori.

Resonance gives rise to librational tori, recall \Fig{Delta01PhaseSpace}. These are most easily seen when one of the components of $\omega$ is constant, such as the $(0,1,0)$ and $(1,0,1)$ resonances labeled in the figure. However, even when both components of $m$ nonzero---such as the $(1,-1,0)$ resonance seen in the figure---the frequencies vary only slightly as the initial conditions cross a resonance so, on the scale of the lower pane of \Fig{FreqMapZoomeps015}, these resonances also appear to be flat intervals. 

There are two intervals in the lower pane of \Fig{FreqMapZoomeps015} where $\omega_T$ appears to vary smoothly. However, this smooth variation of $\omega$ is interspersed with resonant intervals, as can be seen in the upper pane, which is an enlargement around $z_0 = 0$. Resonances can be systematically identified by looking for cases in which $|m \cdot \omega -n| < c$, for some small $c$. The figure identifies all resonances to order 12 that occur for more than one consecutive $z_0$ value, using $c = 10^{-8}$. Note that the vertical scales of $\omega_1$ (shown on the left) and $\omega_2$ (on the right) are different for the upper pane. On the scale of this figure, the component $\omega_1$ is still appears to be constant across a resonance; however, the variation of $\omega_2$ is now more visible since it is displayed on a smaller scale. Since $\omega_2$ varies and $m \cdot \omega = n$ in the resonance, $\omega_1$ must vary as well.

\InsertFig{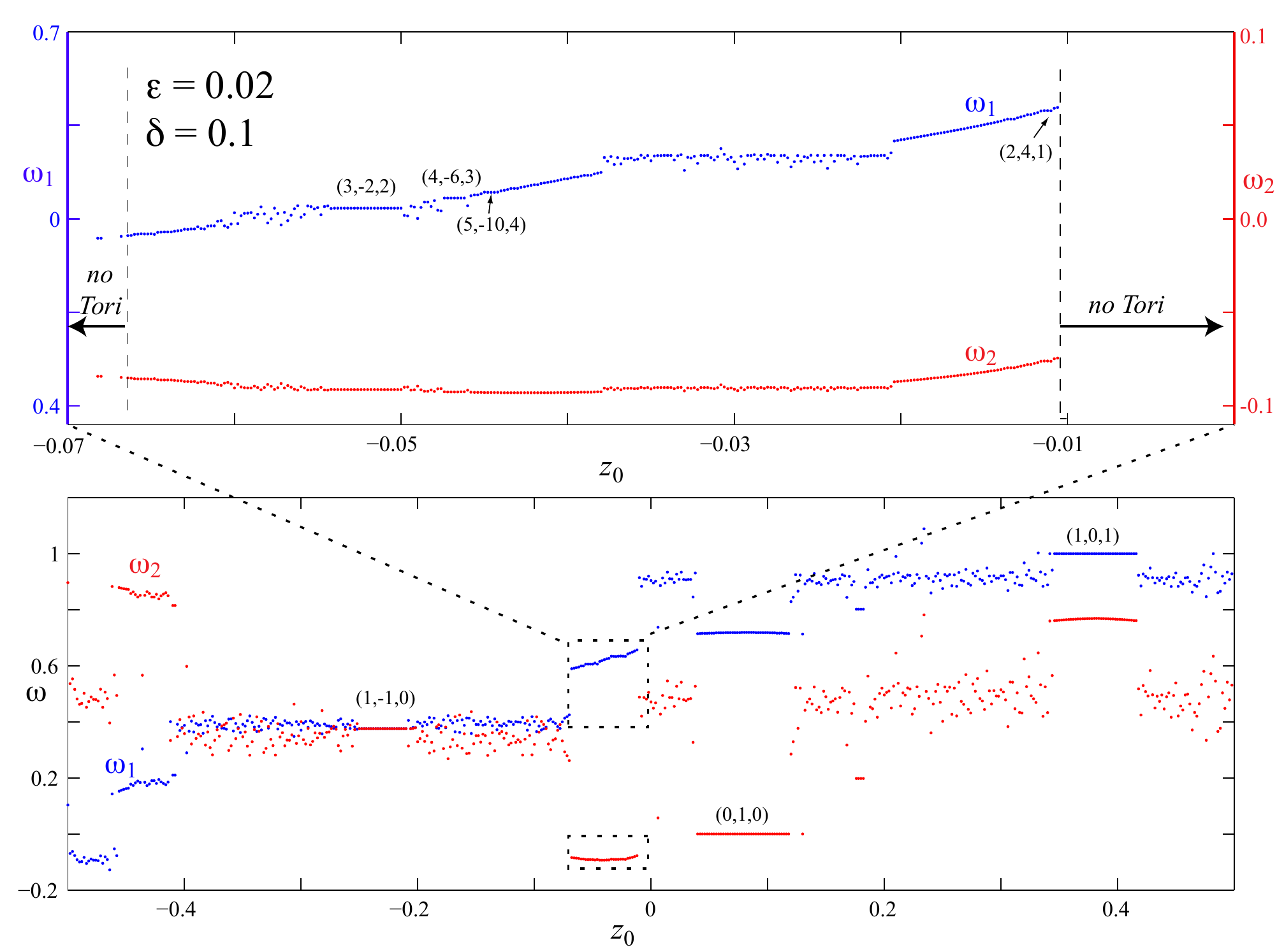}{Numerically computed frequencies, $\omega_T$, as a function of $z_0$ for $\eps = 0.02$ and $\delta = 0.1$ with $x_0 = (0,0)$ and $T=10^7$ (lower pane) and $T=10^8$ (upper pane) for the frequency map \Eq{periodicOmega}. There appear to be invariant tori only in the range $z_0 \in(-0.0663,-0.0104)$, and the upper plot is an enlargement of this region. Resonances up to order $12$ are labeled.}{FreqMapZoomeps02}{6in}

Though there were two small intervals in \Fig{FreqMapZoomeps015} containing invariant tori, \Fig{FreqMapZoomeps02} suggests that when $\eps  = 0.02$ there are rotational tori only in a small interval around  $z = -0.05$. The upper pane is an enlargement of this region. Recall from \Fig{TorusBounds}, there are no tori outside the interval $(z_L,z_U)= (-0.0663, -0.0104)$; this range is also indicated in the figure by the dashed vertical lines. The endpoints of the region of smooth variation of $\omega$ clearly correlate with the interval $(z_L,z_U)$, confirming that smooth variation of $\omega$ indicates the existence of tori. As noted earlier, there are chaotic orbits trapped by tori within $(z_L,z_U)$.

\InsertFig{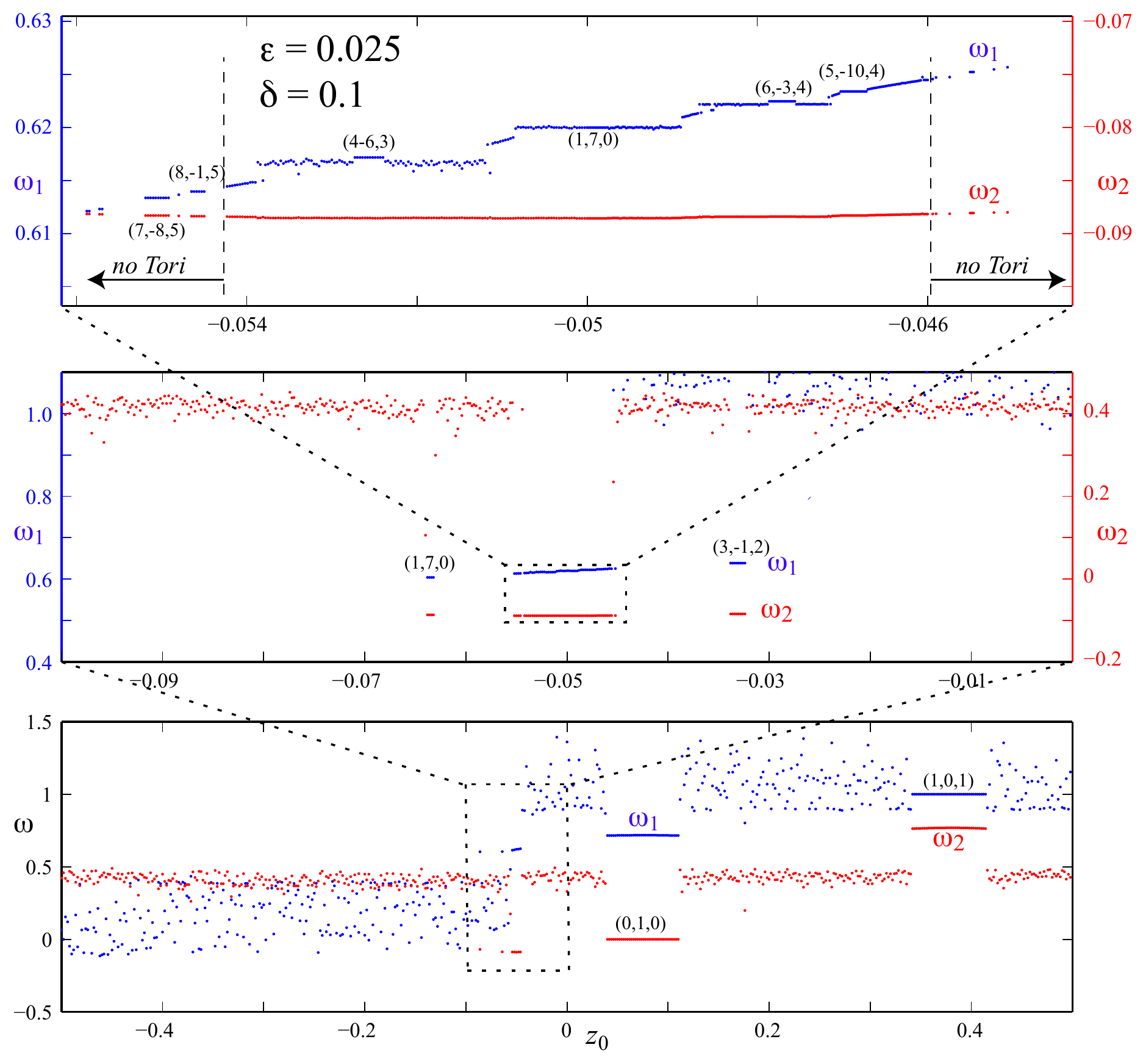}{Numerically computed frequencies, $\omega_T$, as a function of $z_0$ for $\eps = 0.025$ and $\delta = 0.1$ with $x_0 = (0,0)$ and $T=10^7$ (lower pane)  $T=10^8$ (middle pane) and $T=10^9$ (upper pane) for the frequency map \Eq{periodicOmega}. There are no invariant tori outside the interval $z = [-0.05426, -0.04598]$}{FreqmapZoomeps025}{6in}

For $\eps = 0.025$,  $(z_L,z_U) = (-0.054260,-0.045980)$, as shown in \Fig{FreqmapZoomeps025}. It is noteworthy that much of this interval is filled by resonant or chaotic orbits; this can be seen in the uppermost pane of this figure. The Cantor set of tori for this value of $\eps$ is sparse indeed.

\InsertFig{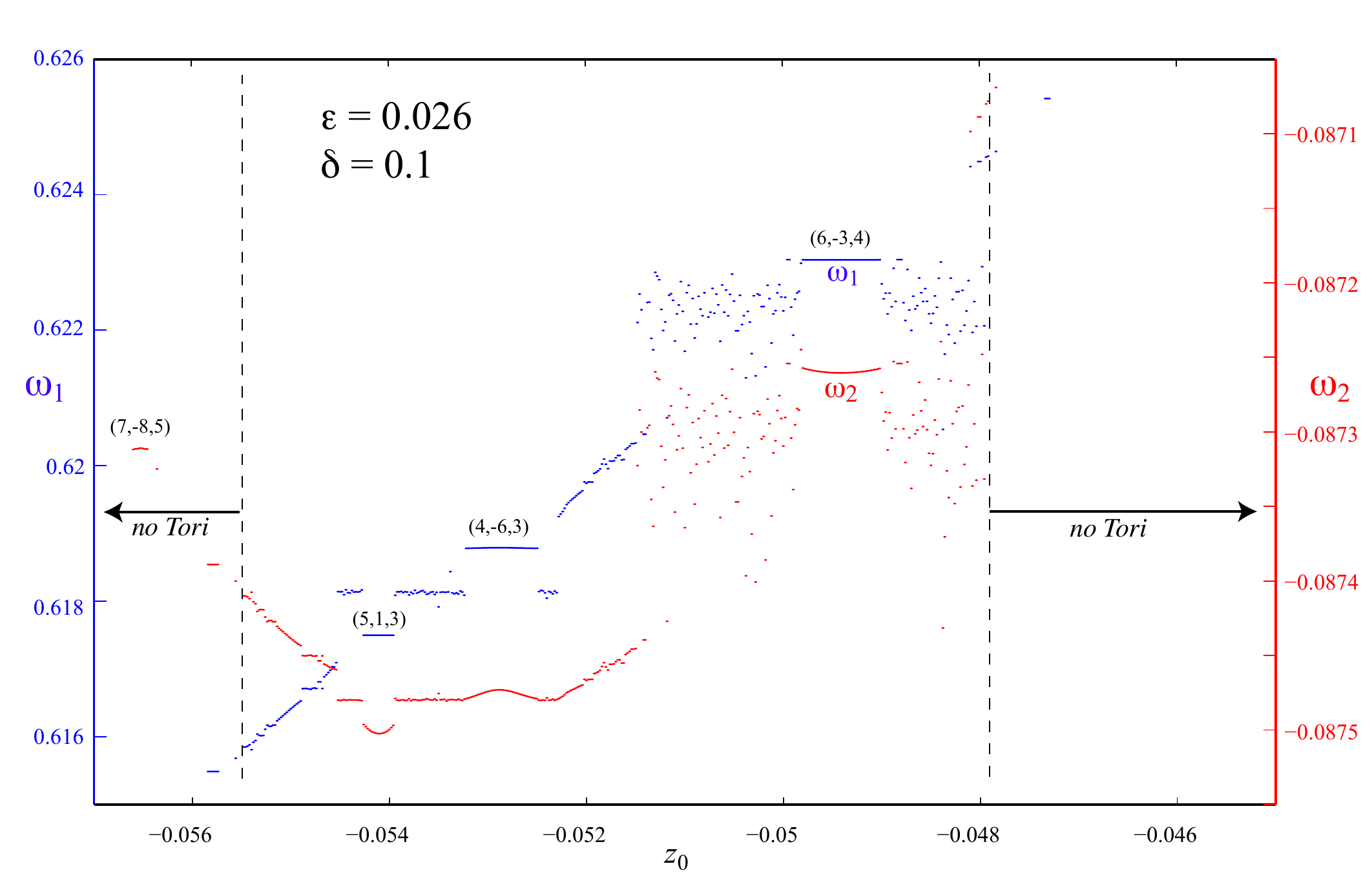}{Numerically computed frequencies, $\omega_T$, as a function of $z_0$ for $\eps = 0.026$ and $\delta = 0.1$ with $x_0 = (0,0)$ and $T=10^9$ for the frequency map \Eq{periodicOmega}.}{FreqmapZoomeps026}{6in}

When $\eps = 0.026$, there are tori only in the interval $(z_L,z_U) = (-0.055500,-0.047900)$. Unlike the previous figures, the upper bound that we found in \Sec{exploration} is not confirmed by the frequency map in \Fig{FreqmapZoomeps026}: the frequencies near this point vary irregularly, at least on the scale of the figure. It is possible; however, that there is a nearly isolated invariant torus at this location.

When $\eps$ is increased to $0.027$ or beyond, it is difficult to compute a meaningful enlargement of the frequency map for $(z_L,z_U)$ as this requires increasing $T$ beyond $10^9$. Nevertheless, the frequency maps confirm the conclusions of the previous methods that $\eps_c \approx 0.0273$.

\section{Discussion}\label{sec:discussion}
We have shown that the break-up of two-dimensional, rotational tori in a three-dimensional, volume preserving map has a number of features similar to those found for the area-preserving case. Namely, the ``last torus" appears to be the limit of a family of resonances. However, we have not been able to discover if there is an analogue of the dynamical self-similarity discovered by John Greene for the standard map.  We also saw that the time for an orbit to cross a region has a power-law singularity just like that discovered by Boris Chirikov for the two-dimensional case. The implication is that destroyed invariant tori may leave remnants like the cantori discovered by Ian Percival for  symplectic maps. However, as of yet, we know of no general construction or even any examples of such invariant cantori sets for the volume-preserving case.

There are many open questions. Birkhoff's theorem implies that the set twist maps with invariant circles is closed since the circles are Lipschitz \cite[\S 5.4]{Herman83a}. Is this true for the volume-preserving case? Can one bound the vertical extent, $\Delta(\cT)$, of rotational tori? Is the set of parameter values for where there are rotational tori itself closed?
Is there an analogue to the robustness of circles with noble rotation numbers to the higher-dimensional case? Since the noble numbers are quadratic irrationals, it has long been speculated that cubic irrationals could fill this role in the two-frequency case.
As Mather demonstrated, it is easy to show that the standard map has no rotational invariant circles for $k > \frac{4}{3}$ \cite{MacKay85}; can one find an explicit ball in parameter space outside of which the family \Eq{perturbed} has no rotational tori?

There are two natural methods for numerical approximation of invariant tori. The first is to look for sequences of periodic orbits whose rotation vectors limit to one for a given torus. The second is to use a Fourier series approximation for the embedding $K:\bT^d \to \bT^d \times \bR^k$ that conjugates the dynamics on the torus to the rigid rotation $\theta \mapsto \theta +\omega$. These complementary methods should lead to more insight into the break-up of tori, and we hope to report on them in the future \cite{Fox11}.


\bibliographystyle{model1-num-names}	

\bibliography{vpTori}
\clearpage

\end{document}